%% file: B2G-22-004_temp.tex
\documentclass[11pt,twoside,a4paper,cmspaper,final,collab]{cms-tdr}
\def\svnVersion{ba1030d}\def\svnDate{2026/04/19}\def\cmsCernNoTag{CERN-EP-2026-021}\def\cmsCernDate{\today}\def\cmsMessage{Submitted to Physics Letters B}
\begin{document}\cmsNoteHeader{B2G-22-004}

\newcommand{\pp}{\ensuremath{\Pp\Pp}\xspace}
\newcommand{\Wb}{\ensuremath{\PW\PQb}\xspace}
\newcommand{\tW}{\ensuremath{\PQt\PW}\xspace}
\newcommand{\Zt}{\ensuremath{\PZ\PQt}\xspace}
\newcommand{\Ht}{\ensuremath{\PH\PQt}\xspace}
\newcommand{\Zbb}{\ensuremath{\PZ\PQb\PAQb}\xspace}
\newcommand{\ee}{\ensuremath{\Pep\Pem}\xspace}
\newcommand{\ST}{\ensuremath{S_{\text{T}}}\xspace}
\newcommand{\mjj}{\ensuremath{m_{\text{jj}}}\xspace}
\newcommand{\mrec}{\ensuremath{m_{\text{rec}}}\xspace}
\newcommand{\kappaW}{\ensuremath{\kappa_\PW}\xspace}
\providecommand{\PQY}{\HepParticle{Y}{}{}\xspace}
\providecommand{\PQT}{\HepParticle{T}{}{}\xspace}
\providecommand{\PQB}{\HepParticle{B}{}{}\xspace}
\providecommand{\PGgst}{\HepParticle{\PGg}{}{\ast}\xspace}
\newcommand{\kappaWY}{\ensuremath{\kappa_\PW^{\PQY}}\xspace}
\newcommand{\kappaWT}{\ensuremath{\kappa_\PW^{\PQT}}\xspace}
\newcommand{\DY}{\ensuremath{\PZ/\PGgst}\xspace}
\newcommand{\ttV}{\ensuremath{\ttbar\PV}\xspace}
\newcommand{\mVLQ}{\ensuremath{m_{\text{VLQ}}}\xspace}
\newcommand{\GammaVLQ}{\ensuremath{\Gamma_{\text{VLQ}}}\xspace}
\newcommand{\WJet}{{\PW + \text{jets}}\xspace}
\newcommand{\ZJet}{{\DY + \text{jets}}\xspace}
\newcommand{\WHF}{{\PW + \text{HF}}\xspace}
\newcommand{\WlF}{{\PW + \text{light-flavor}}\xspace}
\newcommand{\DRljc}{\ensuremath{\Delta R(\Pell, \text{closest jet})}\xspace}
\newcommand{\DRljL}{\ensuremath{\Delta R(\Pell, \text{leading jet})}\xspace}
\newcommand{\DRjj}{\ensuremath{\Delta R(\text{leading jet},\text{closest jet})}\xspace}
\newcommand{\DPljL}{\ensuremath{\Delta \phi (\Pell, \text{leading jet})}\xspace}
\newcommand{\DPlmet}{\ensuremath{\Delta \phi (\Pell, \ptvecmiss)}\xspace}
\newcommand{\DPjmet}{\ensuremath{\Delta \phi (\text{leading jet}, \ptvecmiss)}\xspace}

\cmsNoteHeader{B2G-22-004}
\title{Search for the single production of vector-like quarks decaying into a \texorpdfstring{\PW}{W} boson and a \texorpdfstring{\PQb}{b} quark using single-lepton final states in proton-proton collisions at \texorpdfstring{$\sqrt{s} = 13\TeV$}{sqrt(s) = 13 TeV}}

\date{\today}

\abstract{
A search is performed for the single production of a heavy vector-like quark (VLQ), decaying into a \PW boson and a \PQb quark. The analysis uses proton-proton collision data collected by the CMS experiment at the CERN LHC at a center-of-mass energy of 13\TeV and corresponding to an integrated luminosity of 138\fbinv. The search targets events with leptonic \PW boson decays. The event signature consists of one electron or muon, large transverse momentum imbalance, at least one jet consistent with coming from the fragmentation of a \PQb quark and having large transverse momentum, and at least one jet in the forward region of the detector. No significant excess over the standard model prediction is observed. Upper limits are set at the 95\% confidence level on the production cross section of a VLQ and its coupling \kappaW to the standard model sector. For a VLQ decaying exclusively into \Wb, the upper limit on \kappaW depends on the VLQ mass and reaches values as low as 0.086 for masses around 1.4\TeV. For $\kappaW = 0.2$ the lower limit on the VLQ mass is 2.4\TeV. These are the most stringent limits to date on the single production of VLQs decaying into \Wb.
}

\hypersetup{
pdfauthor={CMS Collaboration},
pdftitle={Search for the single production of vector-like quarks decaying into a W boson and a b quark using  single-lepton final states in proton-proton collisions at sqrt(s) = 13 TeV},
pdfsubject={CMS},
pdfkeywords={CMS, vector-like quarks}}

\maketitle 

\section{Introduction}
\label{sec:intro}

Vector-like quarks (VLQs) are hypothetical colored fermions whose left- and right-chiral 
components transform 
identically under the standard model (SM) gauge symmetry group. They
appear in many extensions of the SM, 
such as models with two Higgs boson doublets~\cite{HDM},
a composite Higgs boson sector~\cite{composite1,composite2},
extra spatial dimensions~\cite{ExtraDim,ExtraDim2}, and some supersymmetric scenarios~\cite{SUSY}.
The VLQs do not acquire mass through a Yukawa coupling to the Higgs boson field, 
and they are not constrained by the single Higgs boson production rates. They can cancel top quark loop corrections to the Higgs boson mass, and thus address the mass hierarchy problem in the SM. 
They can stabilize the Higgs boson vacuum~\cite{stability,stability1} and
play a role in the electroweak (EW) phase transition, baryogenesis~\cite{EWT},
and gauge coupling unification~\cite{unification}.
 
In general, VLQs can be added to the SM in a variety of weak-isospin multiplets~\cite{handbook, Lyon}, 
having quantum numbers that are either SM-like or exotic. 
They can be produced in proton-proton (\pp) collisions at the LHC in pairs via their strong coupling, 
with the production cross section quickly falling as a function of the VLQ mass,
or singly via EW couplings to the gauge bosons by effectively mixing with the SM quarks.
The single-VLQ production rate is proportional to the square of the EW coupling,
and this process can become the dominant production mechanism at large VLQ masses.

Recent LHC searches for pair-produced VLQs set stringent lower limits on their masses~\cite{xanda, ATLAS-result-1, ATLAS-result-2}, 
but are insensitive to the VLQ EW couplings.
The couplings can be probed by searches for single-VLQ EW production. 
The corresponding limits are only now beginning to reach a sensitivity to coupling values not excluded by indirect constraints, 
such as limits from flavor-changing neutral current decays and EW precision measurements~\cite{handbook,xanda,ATLAS-comb}.
The results of the search reported here probe, for the first time, the VLQ EW coupling below the exclusion limits from 
the EW precision observables over the entire considered VLQ mass range.

The search targets the single production of a heavy VLQ that decays into a \PW boson and a \PQb quark, using the data set collected in 2016--2018 with the CMS detector at the CERN LHC and corresponding to an integrated luminosity of 138\fbinv. 
The VLQ can have either the same electric charge as the SM top quark ($+2/3$) or an exotic charge of $-4/3$, and is
labeled as a \PQT or  \PQY quark, respectively.  
The leading-order (LO) Feynman diagram for this single-production mechanism is shown in Fig.~\ref{fig:fey_diagram}. 
The VLQ is produced via \Wb fusion in association with a \PQb quark and a light-flavor quark that is expected to fragment into a jet in the forward region of the detector. 

\ifthenelse{\boolean{cms@external}}
{
\begin{figure}[htp]
\begin{center}
\includegraphics[width=0.45\textwidth]{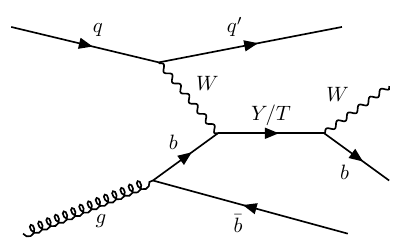}
\caption{Leading-order Feynman diagram for singly produced  \PQY or \PQT quarks. }
\label{fig:fey_diagram}
\end{center}
\end{figure}
}
{
\begin{figure}[htp]
\begin{center}
\includegraphics[width=0.65\textwidth]{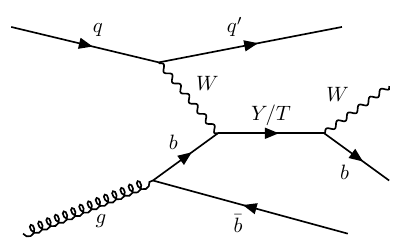}
\caption{Leading-order Feynman diagram for singly produced  \PQY or \PQT quarks. }
\label{fig:fey_diagram}
\end{center}
\end{figure}
}
The vector-like  \PQY and \PQT quarks can belong to different weak-isospin multiplets, 
such as singlet \PQT, doublet (\PQT, \PQB) or (\PQB,  \PQY), or triplet (\PQT, \PQB, \PQY), 
where \PQB is a VLQ with the same electric charge as the SM \PQb quark ($-1/3$).
These are referred to as isospin multiplets of dimension 1, 2, and 3, respectively.
In the absence of other non-SM low-mass states, the  \PQY quark, because of its charge, 
can only decay into \Wb, while the \PQT quark can also decay into \Zt and \Ht, depending on the model,
where \PH is the SM Higgs boson.  
If the \PQT quark forms a singlet, its branching fraction converges to $\mathcal{B} (\PQT\to\Wb) = 50\% $ 
in the high-mass limit, based on the Goldstone boson equivalence theorem~\cite{new_vlq_xsec,goldstone}.
If the \PQT quark belongs to a doublet, the branching fraction $\mathcal{B}(\PQT\to \Wb)$ can be as high as 100\%, depending 
on the mixing parameters with the up- and down-type quark sectors.
The coupling strengths of the  \PQY and \PQT quarks to the \PW boson can be characterized by the parameter \kappaW,
whose allowed range depends on the VLQ multiplet dimension. 
For multiplet dimensions 1 and 2, \kappaW can range between 0 and $\sqrt{2}$, while for dimension 3 it can vary between 0 and 2.
In all cases, the maximum value of \kappaW corresponds to a coupling of full EW strength.

The VLQ multiplet dimension also determines the relative contribution of the left- and right-handed components to the mixing with the SM sector. The left-handed component dominates for singlets and triplets, and the right-handed for doublets. By mixing with the SM sector, the VLQs can affect the values of the EW observables.
For example, by mixing with the SM sector, the (B, Y) doublet would introduce corrections to the \Zbb coupling.  
This would reduce the current tension of 2.4 standard deviations in the EW fit~\cite{global_fit, global_fit2} 
from the \PZ boson $A_{\text{FB}}^{\text{b}}$ measurement by increasing the right-handed coupling. In Ref.~\cite{handbook}, it is shown that 
the global EW fit is improved by adding the (B, Y) doublet to the SM, with the best fit \kappaW value of either 0.17 or 0.23, 
depending on the fit assumptions. The search presented here is sensitive to these \kappaW values over a large VLQ mass range. 

The VLQ natural width \GammaVLQ is proportional to the square of \kappaW, 
and increases as the third power of the VLQ mass \mVLQ.
For a VLQ decaying exclusively into \Wb, the width is negligible compared
to the experimental invariant mass resolution for \kappaW values below 0.25 
over the entire VLQ mass range considered, from 0.7 to 2.4\TeV. 
Thus, the analysis is performed in the narrow-width approximation  
($\GammaVLQ \ll \mVLQ$), 
where interference effects with SM processes can be neglected.

The previous CMS search for $\text{VLQ} \to \Wb$~\cite{CMS-YWb} was performed at $\sqrt{s} = 13\TeV$ 
with an integrated luminosity of 2.3\fbinv, and excluded \kappaW values as low 
as 0.35 and 0.5 for the \PQY and singlet \PQT quarks, respectively.
The most recent ATLAS searches~\cite{ATLAS-YWb, ATLAS-YWb-allhad, ATLAS-YWb-latest} 
set 95\% confidence level (\CL) limits on \kappaW, ranging from 0.14 to 0.46 for the \PQY quark and 
from 0.22 to 0.52 for the singlet \PQT quark, depending on \mVLQ.

The analysis presented here uses the CMS \pp collision data set at $\sqrt{s} = 13\TeV$ 
from 2016--2018.  The search is carried out using events with large missing transverse momentum (\ptvecmiss) defined below, 
exactly one identified electron or muon, at least one jet coming from the fragmentation 
of a \PQb quark (labeled a \PQb-tagged jet) with large transverse momentum \pt, 
and at least one jet in the forward region of the detector. 
The analysis employs multivariate techniques to improve sensitivity.

\section{The CMS detector, trigger,  and event reconstruction}
\label{sec:cms}

The central feature of the CMS detector is its 6\unit{m} internal diameter superconducting solenoid
magnet, providing a magnetic field of 3.8\unit{T}. Within the field volume are a silicon pixel and strip tracker, a lead tungstate crystal electromagnetic calorimeter (ECAL), and a brass and scintillator hadron calorimeter (HCAL).
Muon detectors based on gas ionization chambers are interleaved with the layers of the steel
flux-return yoke located outside the solenoid. 
The pseudorapidity range covered by the tracking system is $\abs{\eta} < 2.5$, the muon detector extends up to
$\abs{\eta} = 2.4$, and the calorimeters cover $\abs{\eta} < 3$. The region $3 < \abs{\eta} < 5$ is
instrumented with steel and quartz fiber forward calorimeters. The hermeticity of the detector up
to large values of $\abs{\eta}$ permits an accurate measurement of the momentum balance transverse to the beam direction.
A more detailed description of the CMS detector, together with a definition of the CMS coordinate system, can be found
in Ref.~\cite{cmsdet-old,cmsdet}. 

Events of interest are selected using a two-tiered trigger system. The first level (L1), composed of
custom hardware processors, uses information from the calorimeters and muon detectors to select events  
at a rate of around 100\unit{kHz} within a fixed latency of 4\mus~\cite{CMS:2020cmk}. The second level,
known as the high-level trigger (HLT), consists of a farm of processors running a version of the full
event reconstruction software optimized for fast processing, and reduces the event rate to a few kHz
before data storage~\cite{CMS:2016ngn,CMS:2024aqx}.

The electron data sample was collected using a trigger
that required at least one isolated electron with $\abs{\eta} < 2.5$ 
and a minimum \pt threshold of 27, 32,or 35\GeV, 
or a photon with the minimum \pt either 175 or 200\GeV, depending on the data-taking period.
The muon data sample was collected using a trigger
that required at least one isolated muon with $\abs{\eta} < 2.4$  
and \pt above either 24 or 27\GeV, depending on the data-taking period.

The analysis employs a particle-flow (PF) algorithm~\cite{PF} for reconstruction and identification 
of physics objects in an event, which are classified into charged or neutral hadrons, electrons, muons, or photons, and makes use of an optimized combination of information from all of the subsystems of the CMS detector. 
Charged particle tracks are used to reconstruct the interaction vertices.
The primary vertex (PV) is taken to be the vertex corresponding to the hardest scattering 
in the event, evaluated using tracking information alone, as described in Ref.~\cite{CMS-TDR-15-02}.

Electron candidates are reconstructed by combining the tracking information with energy deposits 
in the ECAL for the range
$\abs{\eta} < 2.5$, excluding $1.44 <  \abs{\eta}< 1.57$, which is the
transition region between the endcap and barrel calorimeters. Tight
identification criteria are applied to select well-reconstructed electron candidates. 
These are identified~\cite{ele} using information on the shower shape, 
track quality, spatial match between the track and the electromagnetic cluster, and the fraction of
total cluster energy in the HCAL. 
The momentum resolution for electrons with $\pt \approx 45\GeV$ from $\PZ \to \Pe \Pe$ decays ranges from 1.6 to 5\%. It is generally better in the barrel region than in the endcaps, and also depends on the bremsstrahlung energy emitted by the electron as it traverses the material in front of the ECAL~\cite{ele,CMS-DP-2020-021}. 

Muons are measured in the pseudorapidity range $\abs{\eta} < 2.4$, with detection planes made using three technologies: drift tubes, cathode strip chambers, and resistive plate chambers. The single muon trigger efficiency exceeds 90\% over the full $\eta$ range, and the efficiency to reconstruct and identify muons is greater than 96\%. Matching muons to tracks measured in the silicon tracker results in a relative transverse momentum resolution, for muons with \pt up to 100\GeV, of 1\% in the barrel and 3\% in the endcaps. Measurements made with cosmic ray muons show that, in the central region of the detector, the \pt resolution is better than 7\% for muons with \pt up to 1\TeV~\cite{muo}. 

Efficiency corrections for lepton trigger, reconstruction, and identification are obtained in bins
of \pt and $\eta$ via the tag-and-probe method~\cite{muo} using data and simulated samples of \PZ boson dileptonic decays, and applied to simulation.

Electrons and muons are required to be isolated to suppress backgrounds from jets containing leptons. 
The relative isolation parameter is defined as the sum of the \pt of charged tracks 
and the calorimeter energy deposits within a cone of 
$\Delta R = \sqrt{\smash[b]{(\Delta\eta)^{2} + (\Delta\phi)^{2}}} = 0.3\,(0.4)$ around the electron (muon) trajectory, excluding the lepton itself, divided by the lepton \pt. Here, 
$\phi$ is the azimuthal angle, and $\eta$ is the pseudorapidity.
The relative isolation parameter is corrected for the effects of additional inelastic \pp collisions 
within the same or adjacent bunch crossings (pileup) and is required to have a value  
$<$0.0354 (0.0646) for electron candidates 
in the central (endcap) calorimeter region, and $<$0.15 for muon candidates.
The isolation
requirement helps to reject jets misidentified as leptons and reduce multijet backgrounds.

Events with exactly one electron or muon candidate having $\pt > 40\GeV$ 
and $\abs{\eta} < 2.1$  are selected for the analysis.
Events with additional leptons having
$\pt > 10\GeV$, $\abs{\eta} < 2.5$, and passing looser isolation
and identification requirements are rejected.

Particles reconstructed by the PF algorithm are clustered into jets with the \textsc{FastJet} package~\cite{fastjet1} 
by using the direction of each particle 
at the PV. Charged hadrons found by the PF algorithm that are associated with pileup vertices 
are not considered. 
Jets are reconstructed using the anti-\kt algorithm~\cite{antiKT} with a
distance parameter of $R = 0.4$ (AK4 jets). An event-by-event jet-area-based correction~\cite{areaCorrection} is applied to remove, 
on a statistical basis, the neutral
pileup contribution that cannot be removed by the charged-hadron subtraction procedure described above. Jet energy corrections are applied to each jet, as a function of \pt and $\eta$, to correct
for the calorimeter response~\cite{jec}.
The corrected jet \pt must be $>$200\GeV for the highest \pt (leading) jet, and $>$30\GeV
for the remaining jets with $\abs{\eta} < 2.4$. 
For the final event selection, at least one jet is required with $\pt > 30\GeV$ and $2.4 < \abs{\eta} < 5.0$.

Jets are identified as \PQb tagged (called \PQb jets), meaning they come from the hadronization of \PQb quarks, 
using the \textsc{DeepJet} algorithm~\cite{DJ}
based on secondary vertex and track impact parameter information. The \PQb-tagged jets are
required to be within the acceptance of the silicon tracker ($\abs{\eta} < 2.4$). We use
both medium and tight working points (WPs) for the algorithm~\cite{BTV-16-002}.
Events with at least one medium WP \PQb-tagged jet are selected.  The algorithm correctly identifies \PQb jets 
with efficiencies of approximately 70\% and 40\% for the medium and tight WPs, respectively.
It misidentifies jets from light-flavor quarks or gluons
(charm quarks) at a rate of about 1\% (20\%) and 0.1\% (2\%) 
for the medium and tight WPs, respectively.
The \PQb tagging efficiency in simulation
is corrected to match the measured efficiency in data as a function of \pt, $\eta$, and the flavor of the jet.

Large-radius jets are used for identifying hadronically decaying top quarks and \PW bosons, 
where individual jets may be merged into a single large-radius jet. 
The large-radius jets are clustered with the \textsc{FastJet} package~\cite{fastjet1} using 
the anti-\kt algorithm~\cite{antiKT} with a distance parameter of $R = 0.8$ (AK8 jets). 
The mass-decorrelated version of the \textsc{DeepAK8} tagger~\cite{deepAK8-MD}, which features a multiclass classifier based on a deep neural network (NN), is used 
to identify hadronic top quark and \PW boson decays.
In the final state of interest, single-VLQ production events are not expected to contain such decays.
Thus, events with identified top quark and \PW boson candidates are attributed to background processes and are vetoed in the final event selection. We use the loose WP, for which light-flavor and gluon jets are misidentified as 
top quark and \PW boson candidates at rates of 2.5 and 5\%, respectively.

The missing transverse momentum vector \ptvecmiss is computed as the negative vector sum of the transverse momenta of all the PF candidates in an event, and its magnitude is denoted as \ptmiss~\cite{CMS:2019ctu}. The \ptvecmiss is modified to account for corrections to the energy scale of the reconstructed jets in the event. 
To suppress multijet and \ZJet events without neutrinos in the final state, 
\ptmiss is required to be greater than 60\GeV.

\section{Simulation of signal and background events}
\label{sec:mc}

Monte Carlo (MC) simulations of signal and SM background events are used to model the data, optimizing selection criteria
for maximum signal sensitivity, and estimate backgrounds. 
Simulations with conditions appropriate for the 2016 data are
generated using the NNPDF 3.0 parton distribution function (PDF) set at LO or
next-to-LO (NLO)~\cite{nnpdf-30}. 
Simulations corresponding to the 2017--2018 conditions are produced using the NNPDF 3.1 PDF set at next-to-NLO (NNLO)~\cite{nnpdf}.

The signal samples are generated at LO with {\MGvATNLO}~2.3.3~\cite{aMCatNLO1,aMCatNLO2}
covering the \mVLQ range from 0.7 to 2.4\TeV, in steps of 0.1\TeV. The \GammaVLQ value is fixed to 10\GeV for all masses, which is small compared to the experimental invariant mass resolution. The signal contributions are normalized using the cross sections from Refs.~\cite{new_vlq_xsec,new_vlq_xsec2}. 

The main backgrounds for this search are inclusive \ttbar production, single top quark production
via \tW and $t$ channels, and production of \WJet. Other smaller contributions include \ttV (\PV = \PZ,\PW), \ZJet, and multijet production.
The inclusive \ttbar and single top quark 
events are simulated at NLO using {\POWHEG}~2.0~\cite{powheg1,powheg2,powheg3}, where showering and hadronization are performed with {\PYTHIA}~8.240~\cite{PYTHIA}.
The \WJet and \ttV events are generated 
at NLO with {\MGvATNLO} also interfaced with {\PYTHIA}. 
The \WJet events are generated in bins of the \PW boson transverse momentum \pt using 
the FxFx merging procedure~\cite{fxfx} for matching jets from the matrix element calculation 
with those from the parton shower.  The  \ZJet and multijet 
events are generated at LO with {\MGvATNLO}~\cite{aMCatNLO1,aMCatNLO2} in bins of \HT, 
where \HT is the scalar sum of the parton transverse momenta in 
the matrix element calculation, using the MLM jet matching procedure~\cite{MLM}.

The MC samples modeling the 2017--2018 data use the underlying event tune CP5~\cite{CP5}, 
while the simulation of the 2016 data employs CUET8PM1~\cite{CP5}, 
except for the \ttbar and single top quark production samples, 
which use the CP5 tune for all data sets.

The background processes are 
normalized to their theoretical cross sections using the highest order calculations available. The \WJet
and \ZJet events are normalized to account for NLO in EW and 
NNLO in quantum chromodynamics (QCD) corrections to the cross sections from Ref.~\cite{wz-nlo}. The cross section for the \ttbar
background is computed at NNLO in perturbative QCD using a soft-gluon resummation at
next-to-next-to-leading-logarithmic (NNLL) precision with the~\textsc{Top++}~2.0 program~\cite{top20}. 
The \ttV processes are normalized to their inclusive cross sections at NLO+NNLL QCD and NLO EW corrections~\cite{ttV}.
For single top quark production, the $t$-channel cross section is calculated with the {\MCFM} program~(v10.1)~\cite{MCFM} 
at NNLO QCD precision~\cite{t-chan}, 
and the \tW-channel cross section is calculated at NLO+NNLL QCD precision~\cite{kidonakis}.

All generated events are processed through the CMS detector simulation based on {\GEANTfour}~\cite{GEANT}. 
Additional minimum-bias events, generated with {\PYTHIA}, are superimposed on the
hard-scattering events to simulate pileup interactions. The simulated events are weighted to reproduce the distribution of the
number of pileup interactions observed in the data.

\section{Event selection and background modeling}
\label{sec:sele}

In the decay of a singly produced heavy VLQ, the \PW
boson and \PQb quark tend to have large momenta pointing in opposite directions.
Thus, events from single-VLQ production are characterized by the presence 
of: one lepton (electron or muon), accompanied by large \ptvecmiss, 
which is mostly aligned with the lepton direction due to the Lorentz boost of the \PW boson; one \PQb jet 
with large 
transverse momentum, primarily in the opposite direction to that of the \PW boson;
and one jet in the forward region of the detector 
from the hadronization of the accompanying light quark.

To optimally reduce the SM background contribution and improve the sensitivity to the VLQ signal, 
an artificial NN is trained using events from MC simulation.
The following 20 input variables are employed for the NN training:

\begin{itemize}
\item{Magnitudes of transverse momenta: lepton \pt, \ptmiss, leading-jet \pt, and \pt of the jet closest to the lepton direction, 
hereafter referred to as the ``closest jet".}
\item{Pseudorapidities: $\eta$ of the lepton, leading jet and closest jet.}
\item{The scalar sums of the transverse momenta: \ST, defined as the sum of the lepton \pt,  \ptmiss, leading-jet \pt,  and \HT, the sum of the AK4 jets \pt.}
\item{Transverse momentum balance of the VLQ decay products, defined as the ratio of the leading-jet \pt  to the \PW boson \pt, where 
the latter is defined as the magnitude of the vector sum of the lepton $\vec{p}_{\text{T}}$ and  \ptvecmiss .}
\item{The \PQb tag discriminant value for the closest jet.} 
\item{Following the definition of $\Delta R$ in Section~\ref{sec:cms},
we use \DRljL, \DRljc, \DRjj, \DPjmet, and \DPlmet.}
\item{Reconstructed invariant mass of a top quark candidate
using the combination of the four-momenta of the lepton and closest jet, and \ptvecmiss, designated as:
$m$(lepton, closest jet, \ptvecmiss).}
\item{The dijet invariant mass of the leading and closest jets, labeled as \mjj(leading jet, closest jet).}
\item{Angular distributions quantifying the aplanarity of the event topology: the angle 
between the \ptvecmiss direction and the plane 
containing the lepton and leading-jet momenta, the angle 
between the leading-jet direction and the plane containing the lepton momentum and the $z$ axis.}
\end{itemize}

The NN is trained on events that pass the baseline event selection requirements described in Section~\ref{sec:cms} and, 
in addition, satisfy $\ST > 500\GeV$ and $\HT > 500\GeV$. 
These event selection criteria are summarized in Table~\ref{tab:selectioncuts1}.

\begin{table*}[htbp]
\centering
\topcaption{Event selection criteria used for NN training.\label{tab:selectioncuts1}}
  \begin{tabular}{ l c}
Object/quantity & Requirement \\
  \hline
Tight isolated \Pe or \PGm    & Exactly 1; $\;\pt > 40\GeV,\;\abs{\eta} < 2.1$  \\
Additional loose leptons     &  None; $\;\pt > 10\GeV,\;\abs{\eta} < 2.5$ \\
Leading jet               & $\pt > 200\GeV,\;\abs{\eta} < 2.4$  \\
Missing momentum                   & $\ptmiss > 60\GeV$ \\ 
Scalar sums &  $\ST > 500\GeV,\;\HT > 500\GeV$ \\ 
  \end{tabular}
\end{table*}

The NN is built using a feed-forward architecture with one hidden layer, and a single output node. 
We use the ReLU activation function for the hidden layer neurons, and the sigmoid function to activate the output node.
The NN is optimized for accuracy using the Adam algorithm~\cite{adam} as a metric and the binary cross-entropy as a loss function.
We validate the modeling of all kinematic variables used in the NN training, 
as well as the resulting NN discriminant distribution, in dedicated background-enriched control regions (CRs) 
and in a pseudo-signal region (PSR). These regions are
described below and summarized in Table~\ref{tab:selectioncuts2}.

\begin{figure*}[bthp]
\centering
\includegraphics[width=0.49\textwidth]{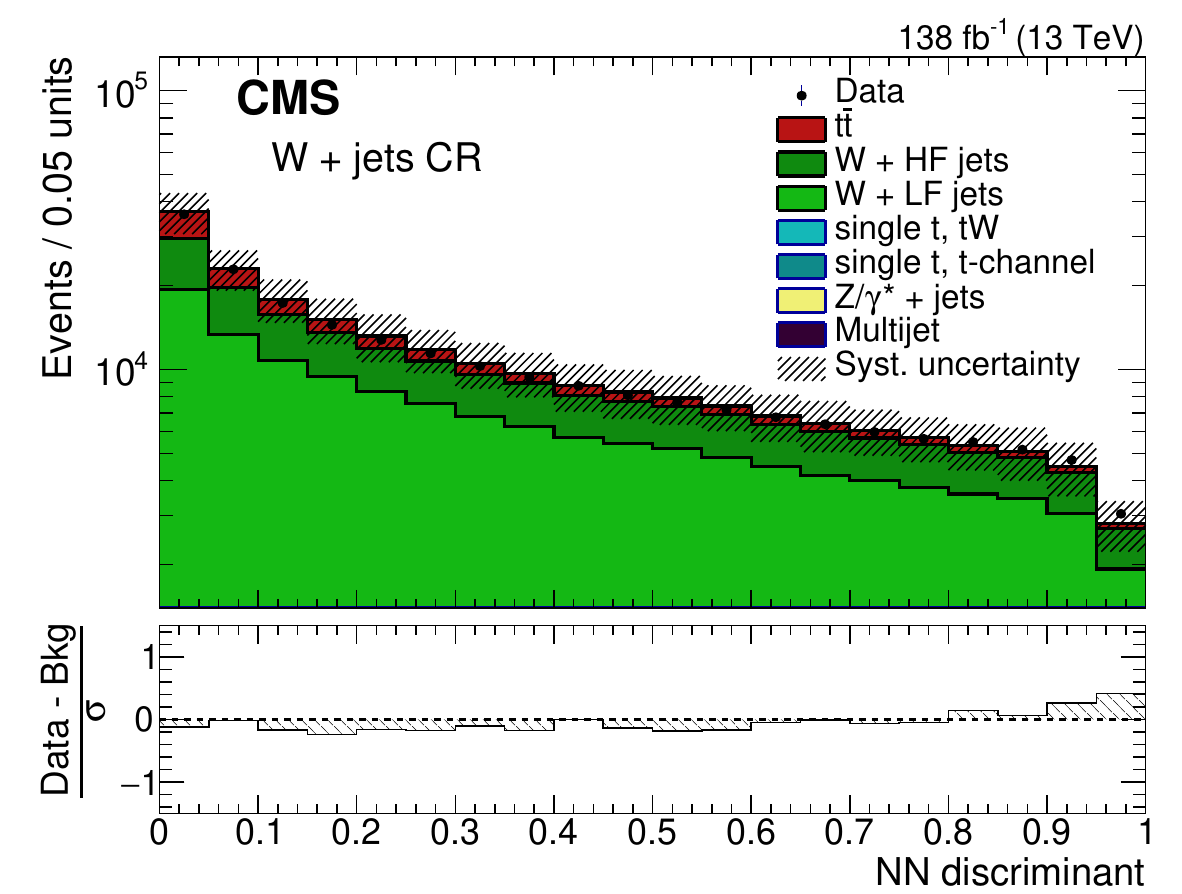}\includegraphics[width=0.49\textwidth]{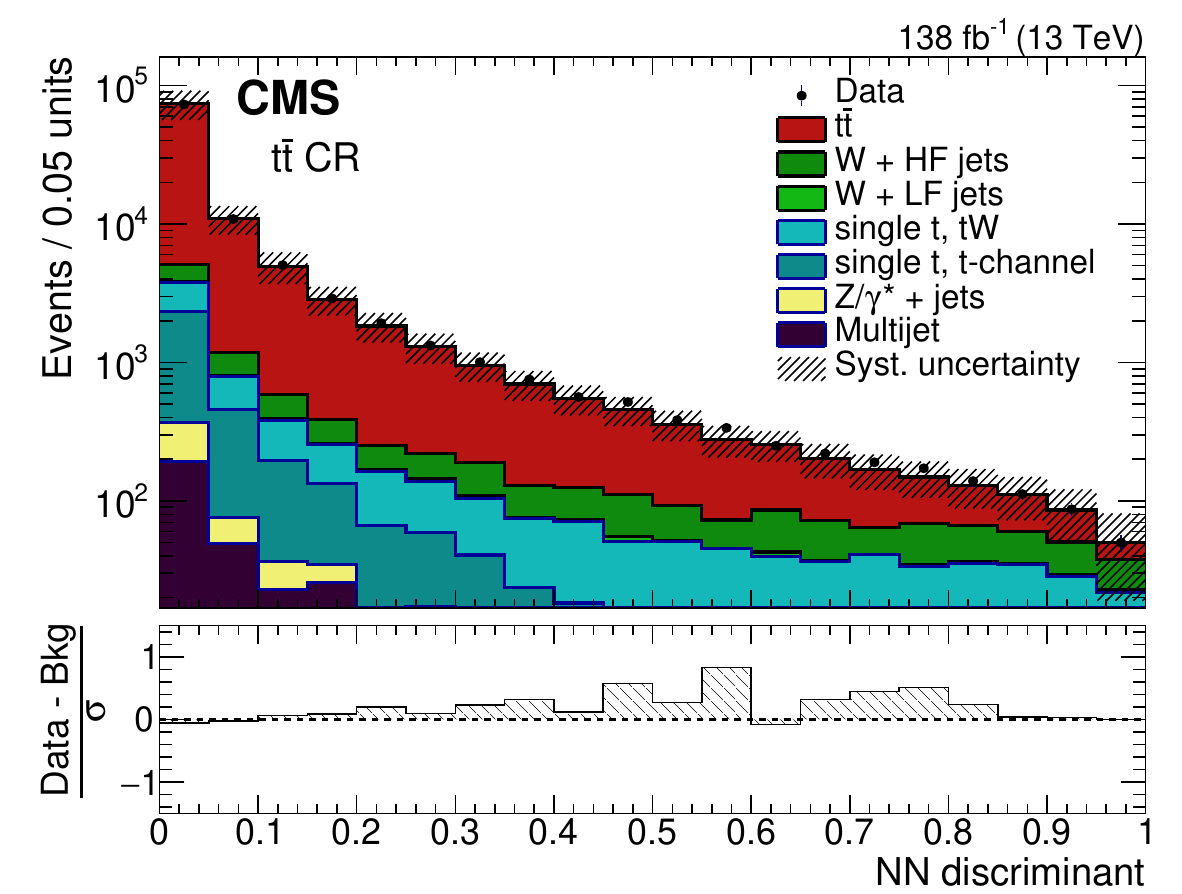} \\
\includegraphics[width=0.49\textwidth]{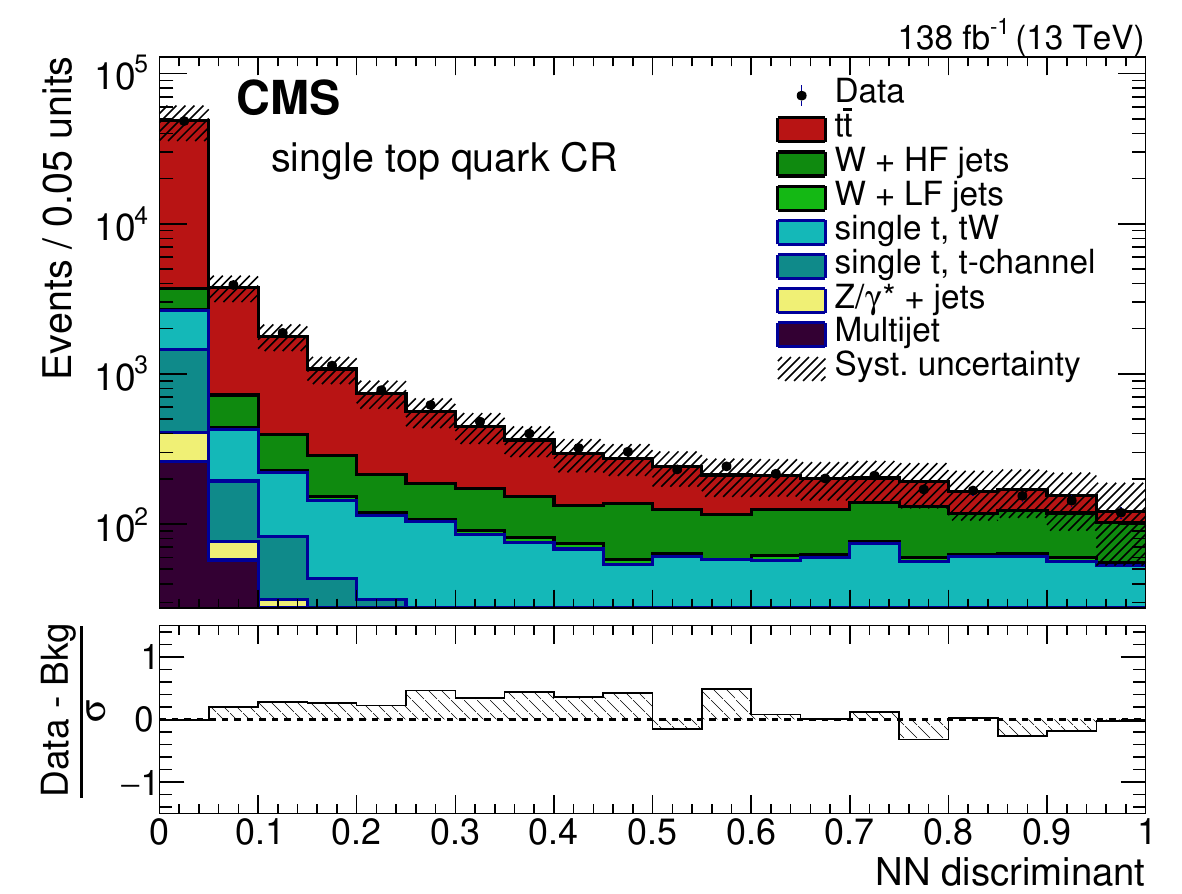}\includegraphics[width=0.49\textwidth]{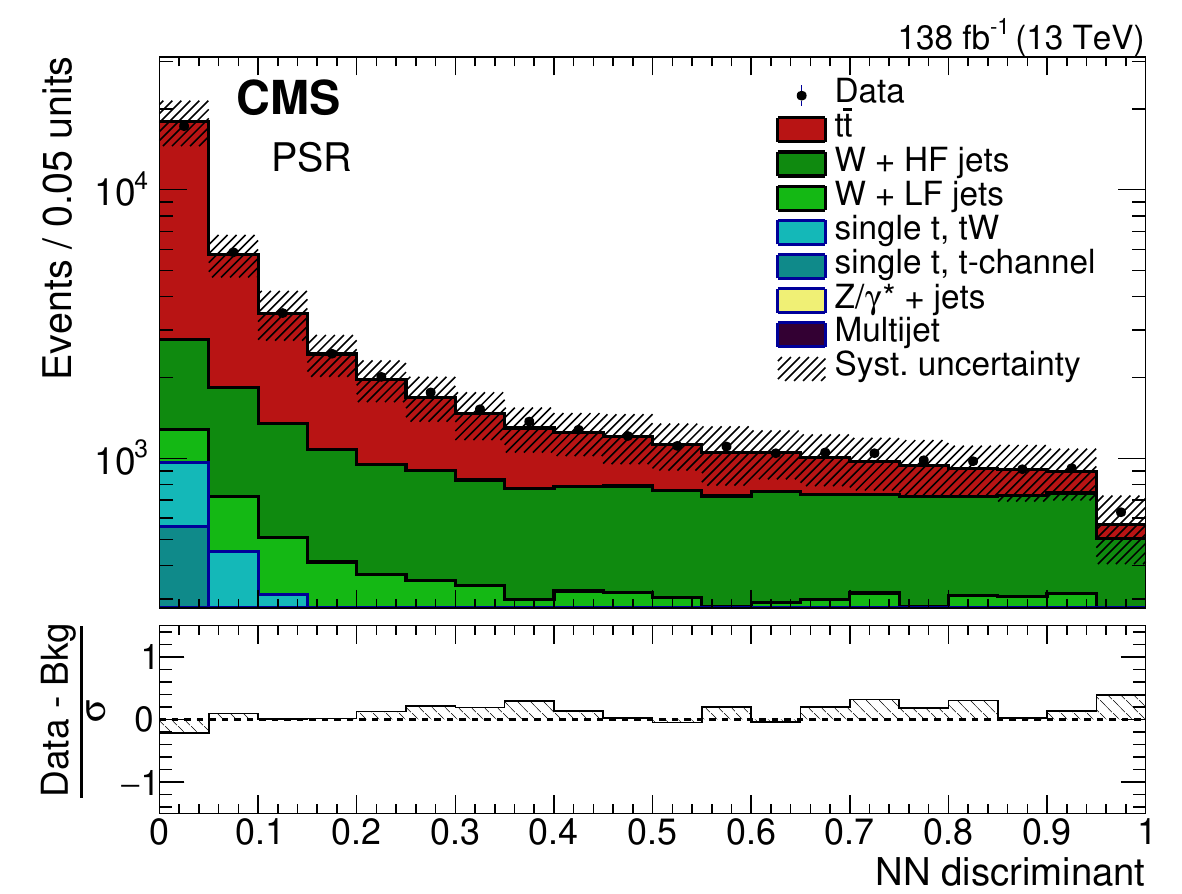}

\caption{Distributions of the NN discriminant in the  \WJet (upper left),  \ttbar (upper right), 
 single top quark (lower left) CRs, and in the 
 PSR (lower right). The data are shown as points and the simulations as colored histograms.  
The vertical bars on the data points represent the statistical uncertainty and the cross-hatched areas give the systematic uncertainty.  
The ``HF" and ``LF" labels refer to heavy- and light-flavor quarks, respectively.  
The lower panels show the differences between the data and the total simulated background predictions, 
divided by the total uncertainty $\sigma$.  }
\label{fig:nnval1}
\end{figure*}

The \WJet CR is defined by applying the following event selection criteria 
in addition to the baseline requirements listed in Table~\ref{tab:selectioncuts1}.
Events with $\ge$1 \PQb jet passing the medium WP are vetoed.
The azimuthal angle difference between
the lepton \Pell and the leading jet must satisfy $\DPljL > 2$.
The azimuthal angle difference between the lepton direction and \ptvecmiss must be 
$\DPlmet < 1$, and the ratio of the \pt of the leading jet to the \pt of the
\PW boson must be less than 2.
The transverse invariant mass of the lepton and \ptvecmiss, defined as
$m_{\text{T}} = \sqrt{\smash[b]{2 p_{\text{T},\Pell} \; \ptmiss \left[ 1 - \cos\Delta\phi(\Pell, \ptvecmiss) \right]}} $,
is required to be greater than 40\GeV to reduce the contribution from multijet processes.  
This selection yields a very pure sample of \WJet events. 
Good agreement is observed between data and MC simulation 
for all kinematic variables used as inputs to the NN in this CR.
The NN discriminant distribution for events in the \WJet CR is shown in Fig.~\ref{fig:nnval1} (upper left).

The modeling of the \WJet process with heavy-flavor (HF) jets originating from
\PQb or \PQc quark fragmentation is validated in a relaxed \WJet CR, in which at most one \PQb jet is allowed in the event. 
In this region, the simulated \WHF contribution is found to be underestimated relative to data by a factor of $1.46 \pm 0.32$, 
although the shapes of all simulated kinematic distributions are in good agreement with data. 
The scale factor (SF) for the \WHF contribution is obtained by performing a maximum likelihood fit of 
the \PQb jet multiplicity distribution in the relaxed \WJet CR to data. 
Both the \WHF component, where one of the jets matches the bottom or charm quark at the generator level, 
and the \WlF (LF) component are left unconstrained in the fit, while other SM processes, 
such as \ttbar and single top quark production, 
are constrained to their theoretical cross section predictions listed in Section~\ref{sec:sysUnc}.
The uncertainty in the SF is evaluated by 
repeating the fit 
with modified event selection criteria, such as removing the \DPljL requirement and 
restricting the event to contain at most four jets. After correcting the \WHF jets contribution, 
good agreement between data and simulation is achieved in all CRs.

To validate the modeling of the \ttbar process, a \ttbar-enriched CR is constructed 
by applying the following event selection criteria 
in addition to the baseline requirements listed in Table~\ref{tab:selectioncuts1}.
Events must have at least two \PQb jets passing the medium WP, 
where the leading jet is not required to have a \PQb tag.
Since \ttbar events are characterized by high jet multiplicity, the lepton is required to have $ \DRljc < 1.5$. 
All other angular requirements are removed to increase the event yield. This selection provides 
a \ttbar CR with a purity exceeding 90\%.
However, the \ST distribution in the \ttbar simulation does not accurately reproduce the shape observed in data. 
To correct for this, a two-parameter linear reweighting function with a negative slope is applied as a function of \ST.
After the correction, good agreement in the modeling of all kinematic variables is obtained.
The NN discriminant distribution in the \ttbar CR is shown in Fig.~\ref{fig:nnval1} (upper right).

A CR with an enhanced single top quark contribution is constructed by modifying the \ttbar CR event selection. The leading jet must pass the medium WP.
The leading jet must have a \PQb tag value above the medium WP.
The \DRljc requirement is removed, 
and a forward-jet veto is applied. 
With these selection requirements, 
the \PW boson \pt spectrum from simulation does not agree with that from data, 
so a correction as a function of the \PW boson \pt is applied. 
Although the single top quark production is not the dominant process in this region, it contributes more than 50\% of the events in the tail 
of the \PW boson \pt spectrum. 
Since the \ttbar modeling is corrected in the dedicated CR in the previous step, this mismodeling is attributed to deficiencies
in the single top quark simulation. 
A two-parameter linear model with a negative slope is used to fit the ratio of data to simulated events in the \PW boson \pt distribution, 
and the result of this
fit is used to correct the modeling of the simulated single top quark events.
The NN discriminant distribution in the single top quark CR is shown in Fig.~\ref{fig:nnval1} (lower left).

The PSR is constructed by applying the following event selection criteria in addition 
to those in Table~\ref{tab:selectioncuts1}.
The angular requirements $\DPljL > 2$, $\DPlmet < 1$, and 
the requirement that $\pt(\text{leading jet}) < 2\, \pt(\PW)$ 
are identical to those used in the \WJet CR selection. In the PSR, the leading jet is
required to have a \PQb tag discriminant value above the medium WP. 
The lepton is required to satisfy $\Delta R > 1.5$ with respect to any other jet with $\pt > 40\GeV$.
In addition, a forward-jet veto is applied
to ensure that events satisfying the final signal region (SR) selection 
remain blinded. The NN discriminant distribution
in the PSR is shown in Fig.~\ref{fig:nnval1} (lower right).

Finally, the SR is constructed from the PSR selection by replacing the forward-jet veto with a requirement of at least one forward jet, 
and by excluding events with hadronically decaying top quark or \PW boson candidates, as well as requiring the NN discriminant to exceed a defined threshold, as described below.

\begin{table*}[htbp]
\centering
\topcaption{Summary of the event selection requirements defining the four CRs, PSR, and SR. 
The NN threshold value $\mathrm{NN}_{\text{thr}}$ for the SR is
optimized separately for each category, as 
described in Section ~\ref{sec:strategy}.\label{tab:selectioncuts2}}
\renewcommand{\arraystretch}{1.3}
  \begin{tabular}{ l cccccc}
Variable & \WJet & \WHF & \ttbar & single \PQt & PSR  & SR  \ \\ 
  \hline
Number of \PQb-tagged jets    & $=$0 & ${\leq}1$ & ${\geq}2$ & ${\geq}2$ & ${\geq}1$ & ${\geq}1$ \\ 
$\DRljc$ & \NA & \NA & $<$1.5 & \NA & $>$1.5 & $>$1.5 \\
$\DPljL$ & $>$2 & $>$2 & \NA & \NA & $>$2 & $>$2  \\ 
$\Delta \phi(\ell, \ptvecmiss )$ & $<$1 & $<$1 & \NA & \NA & $<$1 & $<$1 \\ 
$\pt(\text{leading jet})/\pt(\PW)$ & $<$2 & $<$2 & \NA & \NA & $<$2 & $<$2 \\
$m_{\text{T}}\,(\GeV)$ & $>$40 & $>$40 & \NA & \NA & \NA & \NA \\ 
Leading jet               & \NA & \NA & \NA & \PQb tagged & \PQb tagged & \PQb tagged \\
Number of forward jets    & \NA & \NA & \NA & $=$0 & $=$0 & ${\geq}1$ \\
Number of \PQt or \PW tags    & \NA & \NA & \NA & \NA & \NA & $=$0 \\ 
NN discriminant    & \NA & \NA & \NA & \NA & \NA & $>$$\mathrm{NN}_{\text{thr}}$ \\ 
  \end{tabular}
\end{table*}

\section{Mass reconstruction and event categorization}
\label{sec:strategy}

The VLQ invariant mass is reconstructed using the recursive jigsaw algorithm~\cite{jigsaw}. 
For the event topology in this search the algorithm corresponds to computing an invariant mass 
of the VLQ decay products (\Pell, \PQb, and missing momentum from the \PGn ) 
in the special reference frame boosted 
along the initial proton beam direction ($z$ axis), where the $z$ component of the visible part of the VLQ momentum is zero.
The visible part of the VLQ momentum is the vector sum of the lepton and \PQb jet momenta.
In the boosted reference frame, the $z$ component of the missing momentum is assumed to be zero.
Thus, transforming back to the laboratory frame, the missing momentum obtains a nonzero $z$ component.
In terms of the laboratory frame quantities, the reconstructed VLQ invariant mass \mrec 
is determined using the expression~\cite{jigsaw}:

\ifthenelse{\boolean{cms@external}}
{
\begin{multline}
m_{\text{rec}} = \\
\sqrt{{ m_{\text{vis}}^2 + 2 \left ( \ptmiss \sqrt{{m_{\text{vis}}^2 + {|{{\vec{p}^{\,\text{vis}}_{\text{T}}} }|}}^2\;} -  {\ptvecmiss}\cdot{\vec{p}^{\,\text{vis}}_{\text{T}}} \right)}},
\end{multline}
}
{
\begin{equation}\label{eq:rjs}
m_{\text{rec}} = \sqrt{{ m_{\text{vis}}^2 + 2 \left ( \ptmiss \sqrt{{m_{\text{vis}}^2 + {|{{\vec{p}^{\,\text{vis}}_{\text{T}}} }|}}^2\;} 
-  {\ptvecmiss}\cdot{\vec{p}^{\,\text{vis}}_\text{{T}}} \right)}},
\end{equation}
}
where $m_{\text{vis}}$ and $\vec{p}^{\,\text{vis}}_{\text{T}}$ are the invariant mass and transverse momentum  
of the combined lepton plus leading \PQb jet system, respectively.

The analysis is performed by separately fitting the observed \mrec distributions for several categories of events in the SR described below.
The optimal NN discriminant threshold values $\mathrm{NN}_{\text{thr}}$
are determined by maximizing a figure of merit, defined as $S/(\sqrt{B}+0.1B)$, 
where $S$ and $B$ are the expected number of signal and background events, respectively, for each event category, 
and the second term in the denominator approximates the systematic uncertainty in the  background predictions.

Three exclusive categories of events are defined based on the \PQb-tagged jet information, 
which are labeled as follows.
The ``2 \PQb jets, medium" category contains events with at least two \PQb-tagged jets using 
the medium WP;
the ``1 \PQb jet, medium" category has events with
exactly one \PQb-tagged jet passing the medium but not the tight WP; 
and the ``1 \PQB jet, tight" category requires events to have exactly one \PQb-tagged jet using the tight WP 
and no additional jets passing the medium WP. The electron and muon events are combined in constructing these event categories.
The SR optimization, using the criterion described above, yields $\mathrm{NN}_{\text{thr}}$ values
of 0.87, 0.81, and 0.93 for these categories, respectively.
Each of the three categories is further split into two, based on the sign of the lepton charge, 
labeled as ``\Pep/\PGmp" and ``\Pem/\PGmm" for positively and negatively charged leptons, respectively, to take
advantage of differences in the \PW boson charge asymmetry between the signal and some of the background processes.
Thus, a  total of six event categories are used.

The event selection criteria used to define the SR significantly reduce the SM background contributions, while retaining a high signal acceptance for the range of VLQ masses considered in the analysis. 
In order to mitigate the limited number of MC events and ensure the robustness of the fit to the observed \mrec distributions,
the MC simulated \mrec background distributions are first fitted to a 5-parameter analytical function $f(x)$, defined as:
\begin{equation}\label{eq:anfit}
f(x) = 
\begin{cases}
N  \re^{ -(x-\mu)^2/{2\sigma^2} },  & x \leq \mu \\
N  \re^{- \, s  \left(\text{ln}{\frac{x}{\mu}}\right)^n} , & x > \mu, \, s > 0, \, n > 1, \\
\end{cases}
\end{equation}
where $N$, $\mu$, $\sigma$, $s$, and $n$ are the fitted function parameters. 
The fits are performed for each of the three dominant background processes: \WJet, \ttbar, 
and single top quark in the six event categories, as well as for variations of these shapes from each individual source 
of shape systematic uncertainty described in the next section. 
To account for other SM sources in these fits,  contributions from multijets and 
\DY events are added to the \WJet sample, and the \ttV contributions
to the inclusive \ttbar sample before fitting.

The fitted shapes are verified to agree with the predicted MC background distributions 
by performing Kolmogorov--Smirnov~\cite{KS-test} and $\chi^2$ tests for each background process, each event category, and each systematic variation, 
obtaining p-values above 0.33 for all, with 98\% of them exceeding 0.8.
An F-test~\cite{f-test} shows that a more complex model offers no significant improvement over the function described by Eq.~(\ref{eq:anfit}).

\section{Systematic uncertainties}
\label{sec:sysUnc}

The sources of systematic uncertainty considered either affect only the signal and background rates, or both the rates and the shapes of the simulated \mrec distributions.
Effects from the same uncertainty source are treated as fully correlated across all event categories. 
Theoretical uncertainties are
treated as fully correlated among different data-taking periods, while experimental uncertainties are in most cases 
treated as uncorrelated between the data recorded in 2016, 2017, and 2018, 
since they are mainly of statistical origin and are based on measurements that are independent across the data-taking periods.
The shape uncertainties are modeled by parametrizing the \mrec distributions with nuisance parameters 
that characterize the associated systematic effects varied 
up and down by one standard deviation.
The simulated \mrec distributions are changed according to variations of the nuisance parameters.

The uncertainty in the integrated luminosity is 1.6\%~\cite{lumi16,lumi17,lumi18}. 
The uncertainties in the SM background rates are taken from the respective CMS cross
section measurements at $\sqrt{s} = 13\TeV$, which correspond to 
3.1, 14.2, and 8.4\% for the \ttbar~\cite{ttbar}, single top quark~\cite{singletop},
and \WJet backgrounds~\cite{Wlf}, respectively. In the latter case, the larger uncertainties in the
\WHF jets measurements from Refs.~\cite{Wbb,Wcc}, scaled according to the relative contribution of the \WHF jets process, are also accounted for. 
An additional rate uncertainty, due to the uncertainties in the \WHF scale factor correction~(22\%), is applied separately to the \WHF jets component.
Additional uncertainties in the SM background rate predictions and shape variations~(14--16\%) arise from 
uncertainties in the renormalization and factorization scales. 
These are determined by varying the scales up and down by factors of 2 in the simulation and taking the largest change as the corresponding systematic uncertainty.
Uncertainties arising from the choice of PDFs~(3--8\%) are taken into account according to the PDF4LHC procedure~\cite{PDF4LHC}. 

Trigger and lepton identification efficiencies in simulations are
corrected as functions of the lepton \pt and $\eta$ using decays of \PZ bosons
to leptons in data.  The associated uncertainties of about 2\%  come from the statistical uncertainty in the data.
The uncertainty related to the modeling of pileup is about 1\%, and is evaluated by varying the total \pp inelastic cross section by $\pm$4.6\% relative 
to its nominal value~\cite{PU}. 
The shape uncertainties also include those in the jet energy
scale (5\%), jet energy resolution (2\%), top quark and \PW boson tagging efficiency (3--6\%), and \PQb tagging efficiency (3--5\%).

To account for the mismodeling in the \ttbar simulation,
a two-sided uncertainty band is derived for the \ST correction procedure.  
One side of the band is obtained by removing the correction, and the other by applying the procedure twice.
The same technique is used to determine the uncertainty associated with the simulated
\PW boson \pt distribution for the single top quark process.
The uncertainties due to these corrections are treated independently.
They increase with larger \ST and \PW boson \pt, thereby widening the uncertainty bands at high values of the NN discriminant and \mrec, resulting in systematic uncertainties of 3--8\% and about 50\% in the SRs for \ttbar and single top quark production, respectively.

The uncertainties from fitting the MC background shapes to the analytical function are estimated by
varying the fit parameters by $\pm$1 standard deviation of their fitted values. 
The uncertainties are treated as correlated between the ``\Pep/\PGmp" and ``\Pem/\PGmm" event subcategories 
for each of the ``1 \PQb jet, medium", ``1 \PQb jet, tight", and ``2 \PQb jets, medium" designations 
in the \WJet and \ttbar backgrounds, and uncorrelated in the
single top quark background.

\section{Statistical inference and results}
\label{sec:limits}

\begin{figure*}[htp]
\centering
\includegraphics[width=0.49\textwidth]{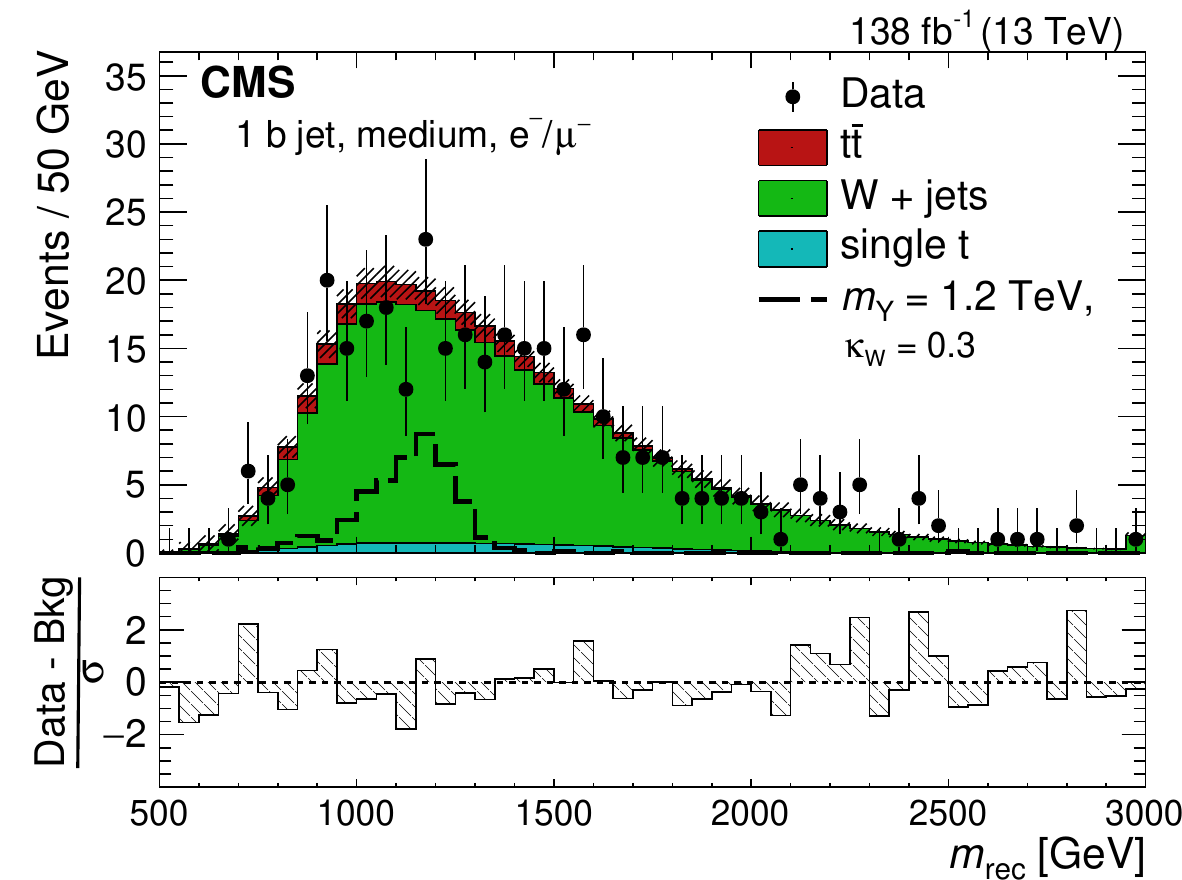}
\includegraphics[width=0.49\textwidth]{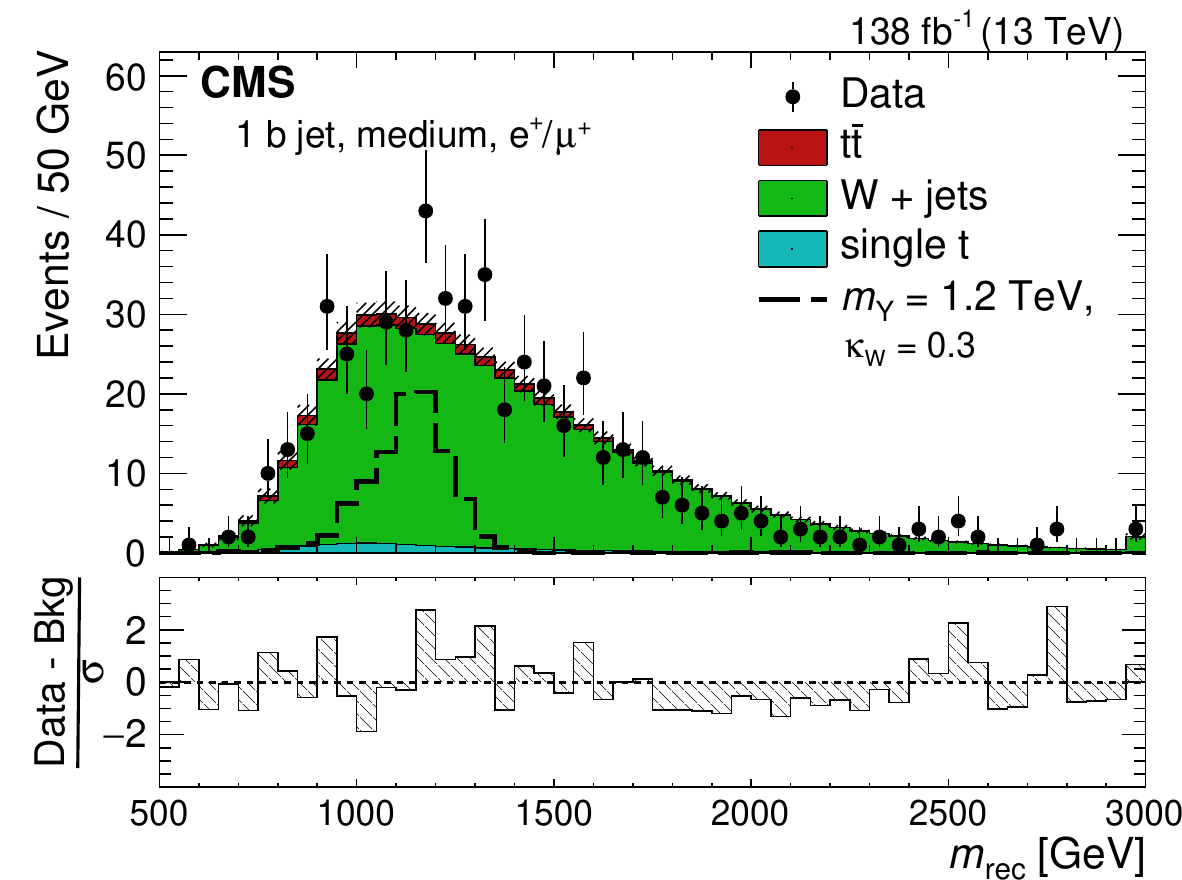}
\includegraphics[width=0.49\textwidth]{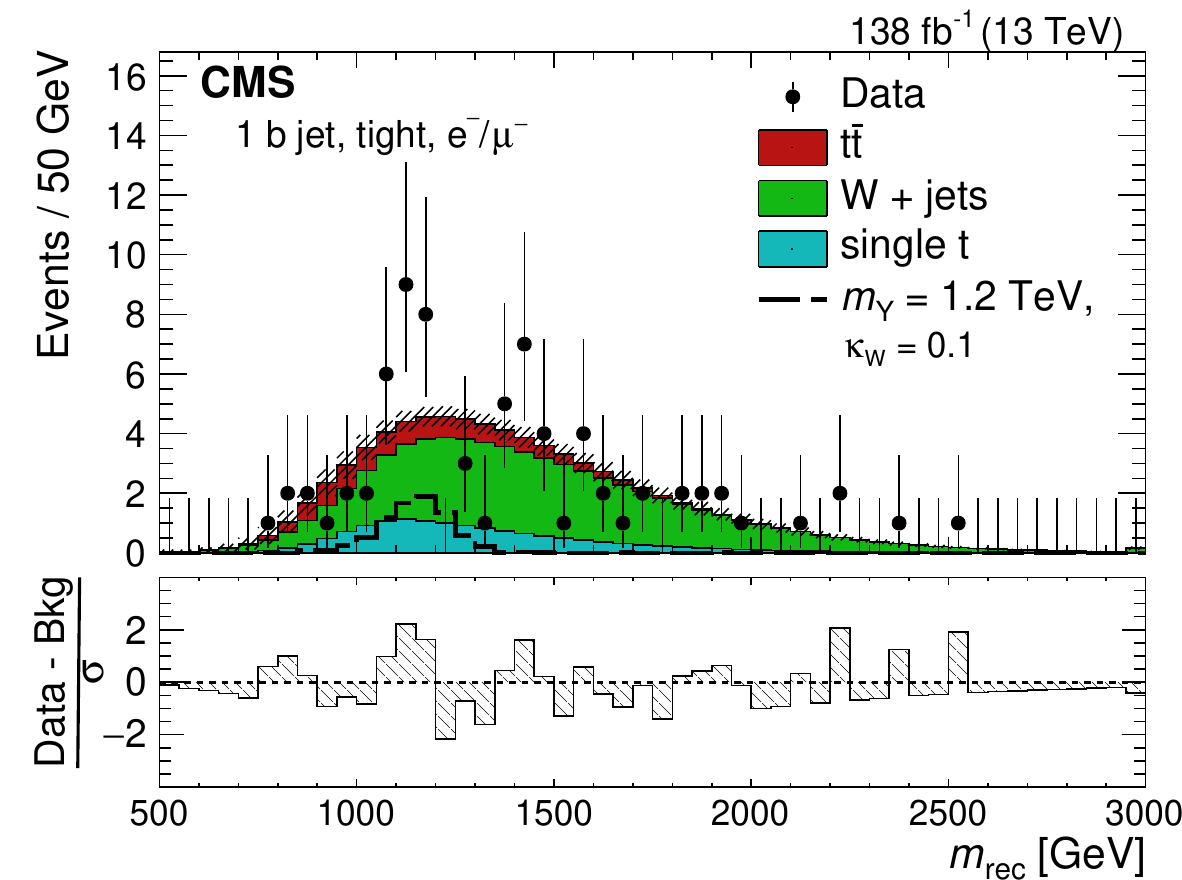}
\includegraphics[width=0.49\textwidth]{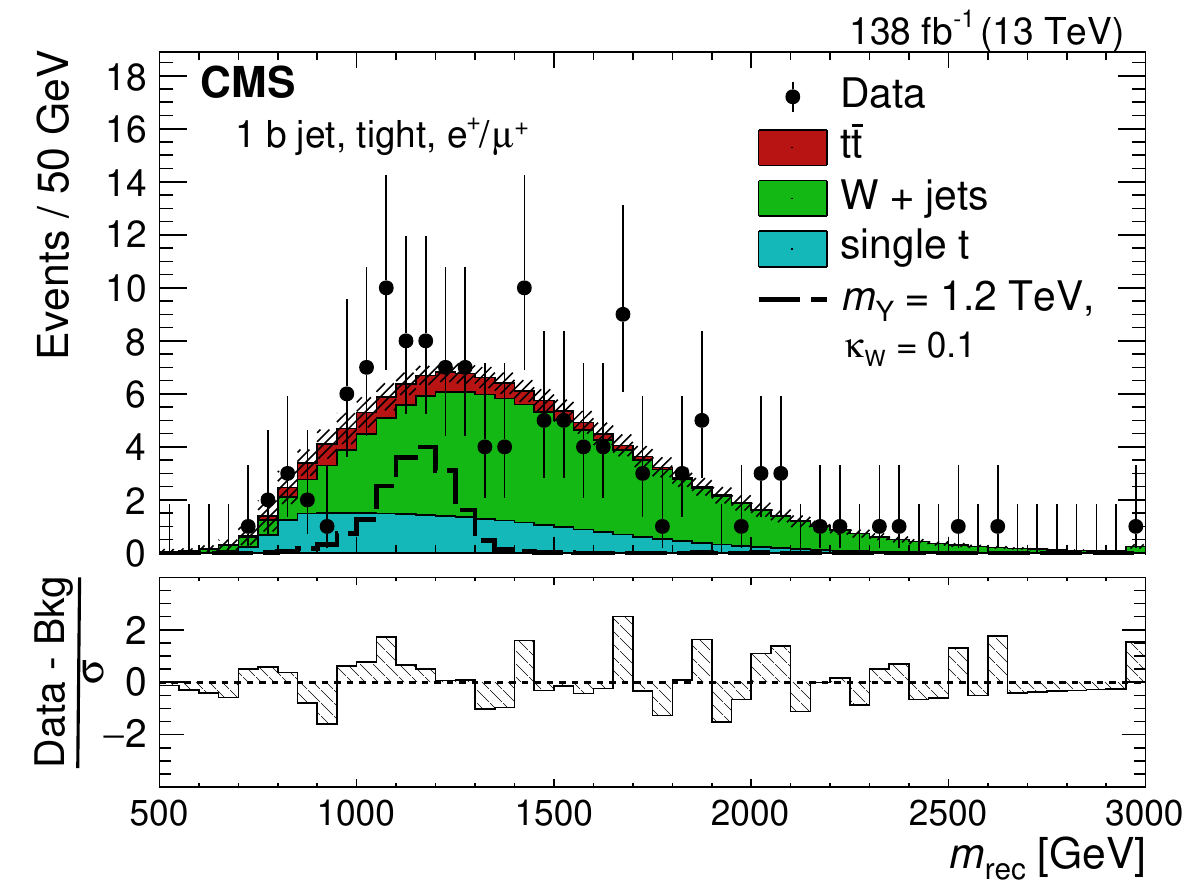}
\includegraphics[width=0.49\textwidth]{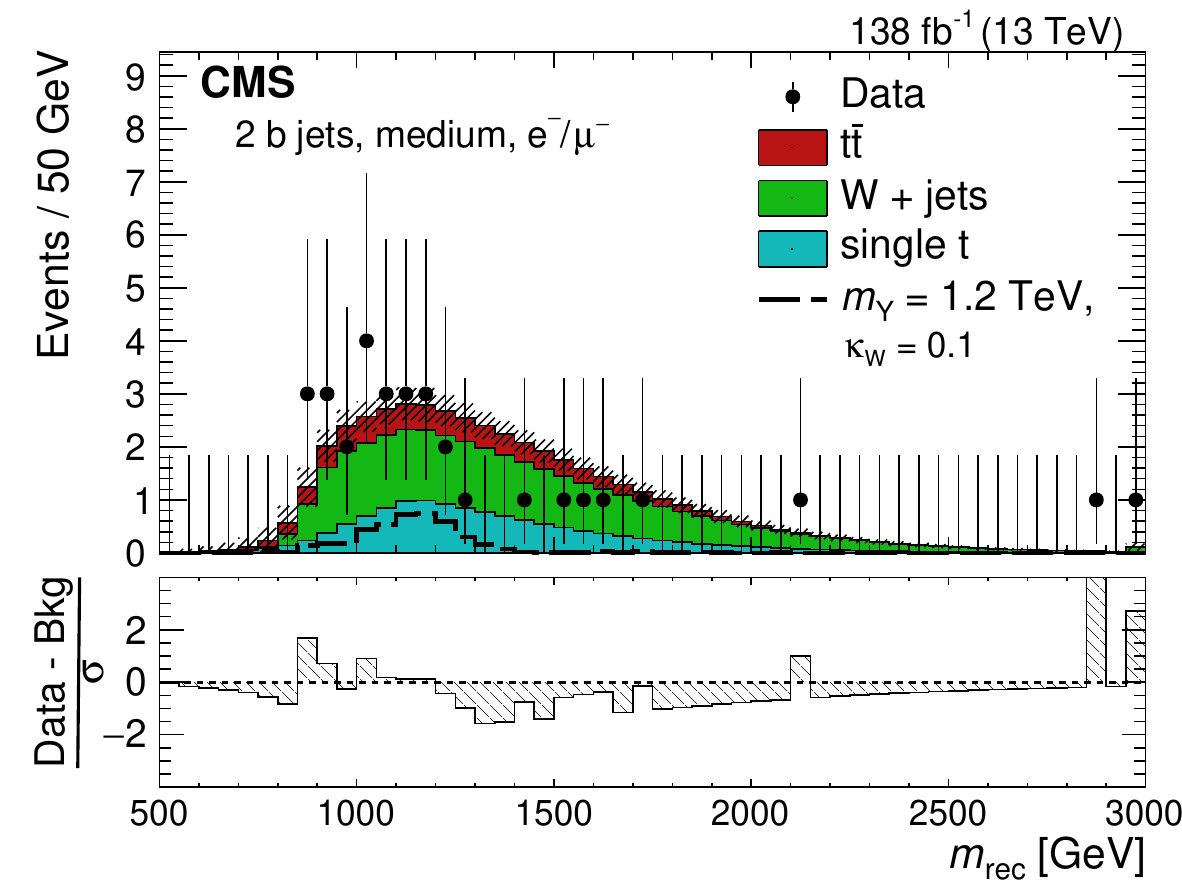}
\includegraphics[width=0.49\textwidth]{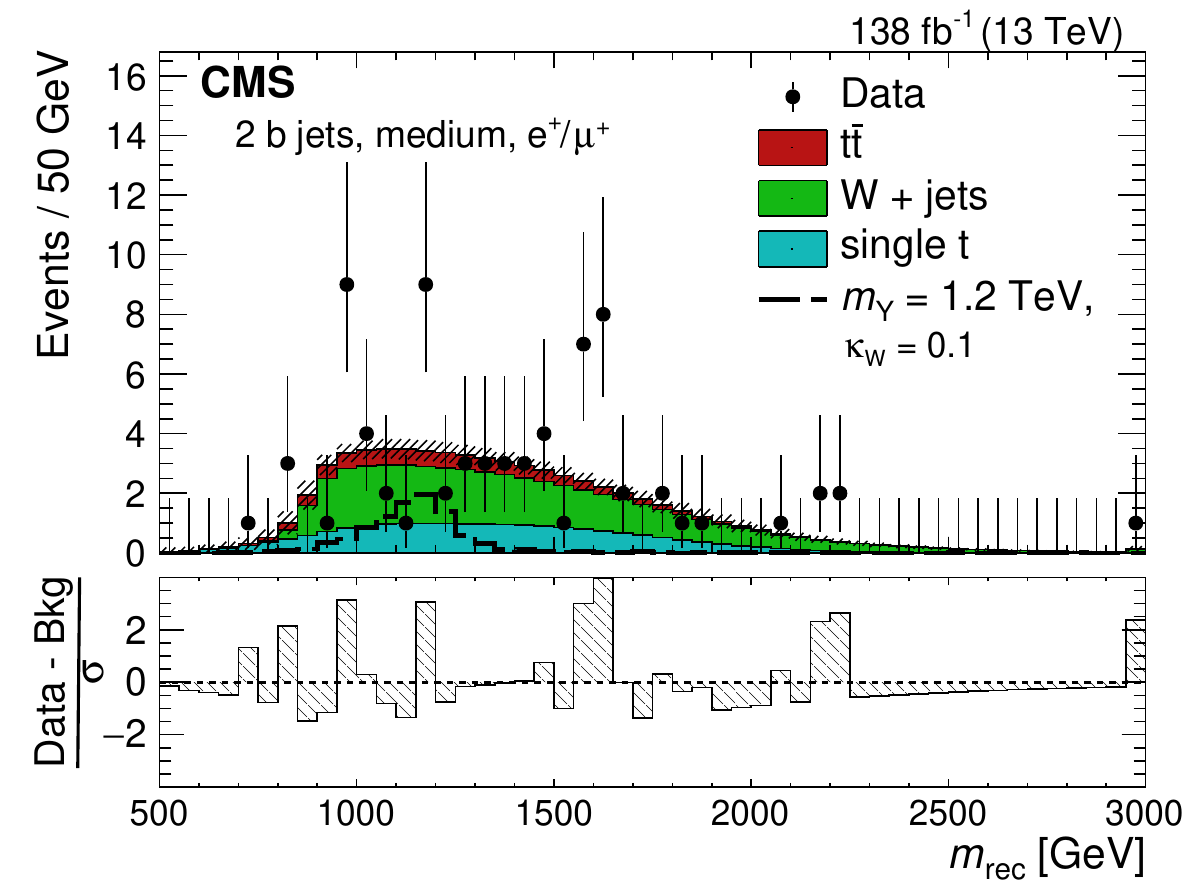}
\caption{The \mrec distributions for each of the six event categories described in the legends 
showing the data (points), the simulated SM backgrounds (colored histograms), 
and the signal predictions for  \PQY quark production (black dashed histogram) 
for a  \PQY quark mass of 1.2\TeV. The coupling parameter \kappaW is set to 0.3 in the 
``1 \PQb jet, medium'' channels, and \kappaW = 0.1 for the others. The vertical bars on the data points represent the statistical
uncertainty in the data.  The hatched area in the upper panels represents the post-fit systematic uncertainties. 
The lower panels show the pulls as hatched bars, defined as the difference between the data and the fitted SM background predictions 
divided by the total uncertainty $\sigma$.}
\label{fig:postfit-mass}
\end{figure*}

Once all CRs have been validated, the SR for each category is unblinded.
The VLQ search is performed in the six SR categories using a simultaneous binned maximum likelihood fit 
of the sum of the simulated SM backgrounds and signal to the observed \mrec distribution.
The SM background \mrec distributions are obtained from the binned shapes fitted 
to the analytical functions given by Eq.~(\ref{eq:anfit}).
Contributions from the SM processes are allowed to vary independently within their systematic
uncertainties, using log-normal priors~\cite{logN2}. 
The nuisance parameters describing the shape uncertainties are constrained using
Gaussian function priors. The shapes of the reconstructed invariant mass distributions for backgrounds 
and signal are parametrized and varied according to the nuisance parameters. 
With this configuration, signal injection tests were performed for several \mVLQ values at various injected signal strengths and did not reveal any significant biases.
The post-fit reconstructed 
mass distributions, with the shape and background normalizations at their maximum
likelihood values, and the systematic uncertainty band corresponding to the post-fit result, 
are shown in Fig.~\ref{fig:postfit-mass}. 

To evaluate the compatibility of the observed data with the SM background prediction, a goodness-of-fit test was performed using the saturated model~\cite{saturated} and yielded a p-value~\cite{pvalue} of 0.12, consistent with the background-only hypothesis. Accordingly, no significant excess of events is observed above the sum of the predicted SM backgrounds.  Upper limits at 95\% \CL on the production
cross section of the $\text{VLQ} \to \Wb$ process are computed with the CMS statistical tool \textsc{Combine}~\cite{combine} using the \CLs criterion~\cite{cls1,cls2,cls3}, where the likelihood is profiled with
respect to the nuisance parameters representing the systematic uncertainties. 
The expected limit is calculated by resampling the pseudo-data
from the background distribution. 

\begin{figure*}[htbp]
\centering
\includegraphics[width=0.49\textwidth]{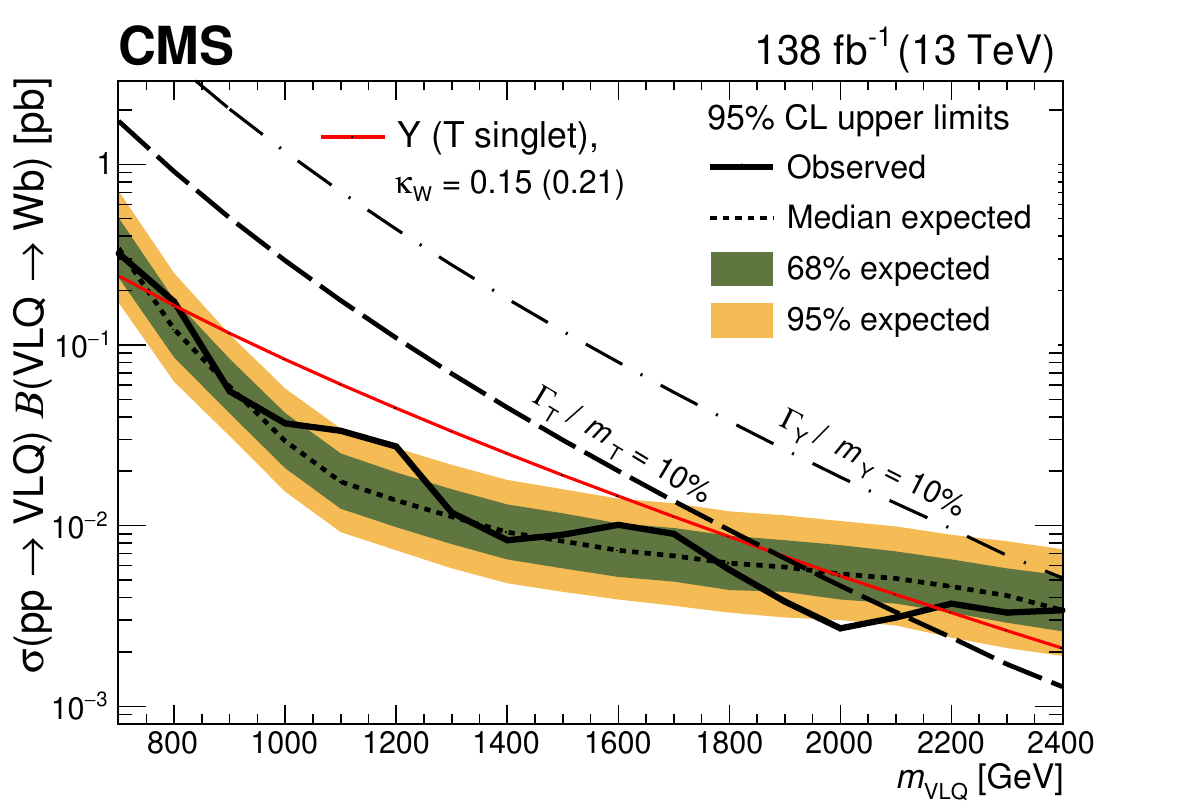}
\includegraphics[width=0.49\textwidth]{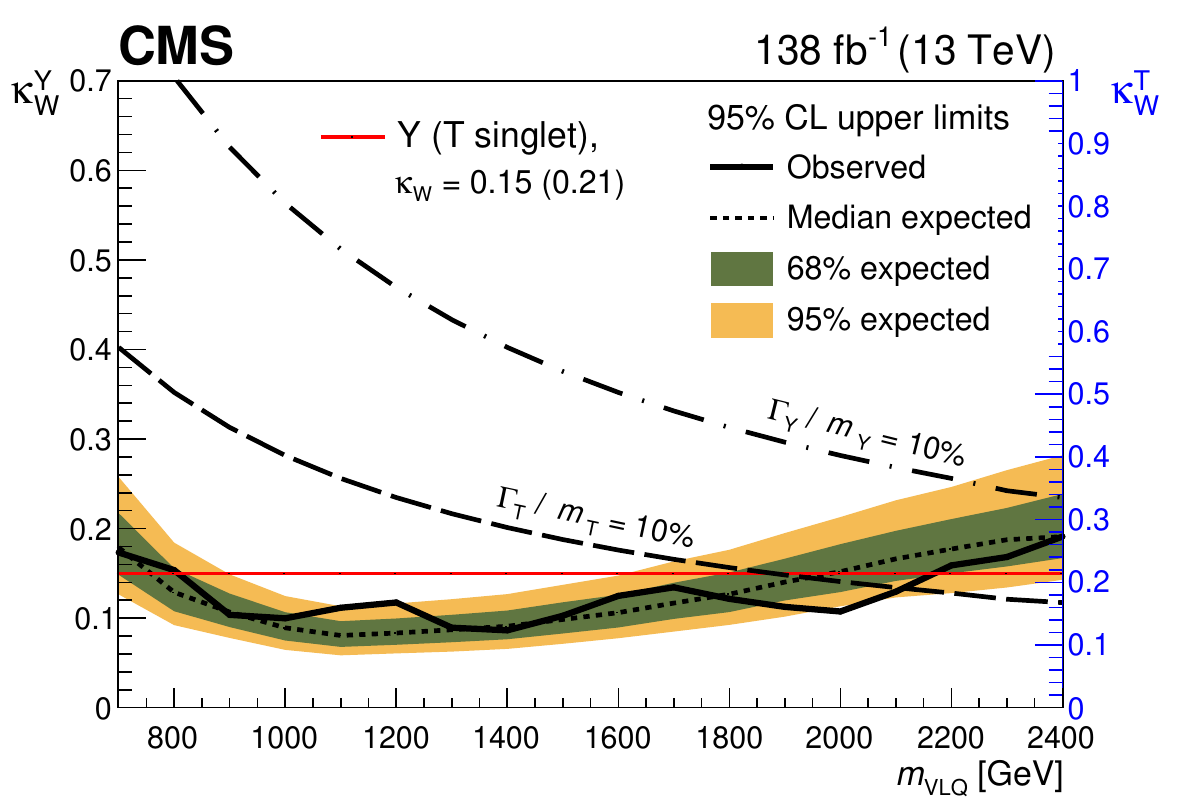}
\caption{
The 95\% \CL upper limits on the VLQ single-production cross section (upper) 
and on the coupling parameter \kappaW (lower) as a function of \mVLQ.  
The solid black curve gives the observed upper limit, while the dotted black curve shows
the expected upper limit, with the green and yellow bands representing the $\pm$1 and $\pm$2 standard
deviation uncertainties in the expected limit, respectively.  
The solid red curve in the upper plot displays the theoretical prediction for the  \PQY and \PQT quark single-production cross section assuming exclusive decay to \Wb and $\kappaW = 0.15$, while the solid red curve
in the lower plot shows the prediction for \kappaWY under the same assumptions. 
These two curves also correspond to the predictions for the \PQT singlet, with 
$\mathcal{B}(\Wb) = 50\%$ and $\kappaW = 0.21$.
In this case, the scale for the prediction of \kappaWT is given by the right vertical axis in the lower
plot. The black dashed and dot-dashed curves show the expected cross sections 
corresponding to a fixed VLQ width to mass ratio $\Gamma/m = 10\%$ 
for the \PQT singlet and  \PQY quarks, respectively. 
Below these curves, the VLQ natural width is smaller than the experimental invariant mass resolution, 
and the narrow-width approximation assumed in the analysis is justified.
}
\label{fig:results}
\end{figure*}

The 95\% \CL expected and observed upper limits on the production cross section and the coupling parameter \kappaW are shown in Fig.~\ref{fig:results}. 
Based on these results, both  \PQY and \PQT VLQs with $\mathcal{B}(\text{Y/T} \to \Wb) = 100\%$ 
and \kappaW values of 0.2 and 0.15, are excluded over \mVLQ ranges of 0.7 to 2.4 and 0.82 to 2.15\TeV, respectively.
Thus, the hypothesis proposed in Ref.~\cite{handbook} 
of adding a (B, Y) doublet to the SM, 
with \kappaW values in the range 0.17--0.23 preferred by the EW fit, is not supported by results of this search.

For the singlet T, the theoretical cross section, selection efficiency, and measured \mrec distributions are
the same as for the  \PQY quark. However, the expected branching fraction $\mathcal{B}(\text{T} \to \Wb)$ is 50\%, half of the 
Y quark value. This means that the singlet \PQT mass exclusion limits are identical to those for the  \PQY quark if
their respective coupling values are scaled as $\kappaWT = \sqrt{2} \kappaWY$.
In addition, the natural width of the singlet \PQT is twice as large as that of the  \PQY quark for the same values of 
\kappaW and mass. Since the upper limits on the \kappaW value are weaker for the singlet \PQT quark, 
and the width increases with VLQ mass, 
the narrow-width approximation used in this search is valid only up to a singlet \PQT quark mass of 2\TeV, 
when the experimental detector resolution of 12\% becomes comparable to the natural width of the singlet \PQT quark.

Tabulated results for Figs.~\ref{fig:postfit-mass} and~\ref{fig:results} are provided in the HEPData record~\cite{hepData}.

\section{Summary}
\label{sec:Summary}

Results have been presented from a search for the single production of the vector-like quarks \PQT and Y, with
electric charges of $+2/3$ and $-4/3$, respectively, and decaying into a \PW boson and a \PQb quark.
The results are based on data collected by the CMS experiment at the CERN LHC from proton-proton
collisions at $\sqrt{s} = 13\TeV$, corresponding to an integrated luminosity of 138\fbinv.

The search is performed using events with large transverse momentum imbalance, exactly one charged lepton, at least one 
jet identified as coming from the fragmentation of a \PQb quark (\PQb tagged), and
at least one jet in the forward region of the detector.
The invariant mass \mrec of the vector-like quark  is
reconstructed with the recursive jigsaw technique using the four-momenta of the charged lepton and
the highest momentum \PQb-tagged jet, combined with the missing transverse momentum.

A profile likelihood fit to the observed \mrec spectrum is performed simultaneously in six different event categories 
based on the number of \PQb-tagged jets in the event and the sign of the lepton charge.  
No evidence for an excess above the expected standard model background due to the
production of a VLQ is observed.
No evidence of an excess due to new physics above the expected background is observed. 
Upper limits at 95\% confidence level are set on the cross sections 
for the single production of vector-like quarks and on their coupling \kappaW to the standard model sector
in the mass range from 0.70 to 2.40 TeV.  

Assuming $\mathcal{B}(\text{Y/T} \to \Wb) = 100\%$, the 95\%
confidence level upper limit on \kappaW depends on the VLQ mass and reaches values as low as 0.086 at VLQ masses around 1.4\TeV. 
For \kappaW values of 0.2 and 0.15, the  \PQY vector-like quark is excluded in the mass range 0.7 to 2.4 and 0.82 to 2.15\TeV, respectively.
For \PQT quarks with $\mathcal{B}(\text{T} \to \Wb) = 50\%$, the upper limits on the production cross section are twice as large, and the \PQT vector-like quark is excluded over the mass range 0.8 to 2\TeV for $\kappaW = 0.2$.
These results represent the most stringent limits to date
on the single production of vector-like  \PQY and \PQT quarks decaying into a \PW boson and a \PQb quark.

\begin{acknowledgments}
We congratulate our colleagues in the CERN accelerator departments for the excellent performance of the LHC and thank the technical and administrative staffs at CERN and at other CMS institutes for their contributions to the success of the CMS effort. In addition, we gratefully acknowledge the computing c enters and personnel of the Worldwide LHC Computing Grid and other c enters for delivering so effectively the computing infrastructure essential to our analyses. Finally, we acknowledge the enduring support for the construction and operation of the LHC, the CMS detector, and the supporting computing infrastructure provided by the following funding agencies: SC (Armenia), BMBWF and FWF (Austria); FNRS and FWO (Belgium); CNPq, CAPES, FAPERJ, FAPERGS, and FAPESP (Brazil); MES and BNSF (Bulgaria); CERN; CAS, MoST, and NSFC (China); MINCIENCIAS (Colombia); MSES and CSF (Croatia); RIF (Cyprus); SENESCYT (Ecuador); ERC PRG and PSG, TARISTU24-TK10 and MoER TK202 (Estonia); Academy of Finland, MEC, and HIP (Finland); CEA and CNRS/IN2P3 (France); SRNSF (Georgia); BMFTR, DFG, and HGF (Germany); GSRI (Greece); MATE and NKFIH (Hungary); DAE and DST (India); IPM (Iran); SFI (Ireland); INFN (Italy); MSIT and NRF (Republic of Korea); MES (Latvia); LMTLT (Lithuania); MOE and UM (Malaysia); BUAP, CINVESTAV, CONACYT, LNS, SEP, and UASLP-FAI (Mexico); MOS (Montenegro); MBIE (New Zealand); PAEC (Pakistan); MSHE, NSC, and NAWA (Poland); FCT (Portugal); MESTD (Serbia); MICIU/AEI and PCTI (Spain); MOSTR (Sri Lanka); Swiss Funding Agencies (Switzerland); MST (Taipei); MHESI (Thailand); TUBITAK and TENMAK (T\"{u}rkiye); NASU (Ukraine); STFC (United Kingdom); DOE and NSF (USA).

\hyphenation{Rachada-pisek} Individuals have received support from the Marie-Curie program and the European Research Council and Horizon 2020 Grant, contract Nos.\ 675440, 724704, 752730, 758316, 765710, 824093, 101115353, 101002207, 101001205, and COST Action CA16108 (European Union); the Leventis Foundation; the Alfred P.\ Sloan Foundation; the Alexander von Humboldt Foundation; the Science Committee, project no. 22rl-037 (Armenia); the Fonds pour la Formation \`a la Recherche dans l'Industrie et dans l'Agriculture (FRIA) and Fonds voor Wetenschappelijk Onderzoek contract No. 1228724N (Belgium); the Beijing Municipal Science \& Technology Commission, No. Z191100007219010, the Fundamental Research Funds for the Central Universities, the Ministry of Science and Technology of China under Grant No. 2023YFA1605804, the Natural Science Foundation of China under Grant No. 12535004, and USTC Research Funds of the Double First-Class Initiative No.\ YD2030002017 (China); the Ministry of Education, Youth and Sports (MEYS) of the Czech Republic; the Shota Rustaveli National Science Foundation, grant FR-22-985 (Georgia); the Deutsche Forschungsgemeinschaft (DFG), among others, under Germany's Excellence Strategy -- EXC 2121 ``Quantum Universe" -- 390833306, and under project number 400140256 - GRK2497; the Hellenic Foundation for Research and Innovation (HFRI), Project Number 2288 (Greece); the Hungarian Academy of Sciences, the New National Excellence Program - \'UNKP, the NKFIH research grants K 131991, K 138136, K 143460, K 143477, K 147557, K 146913, K 146914, K 147048, TKP2021-NKTA-64, and 2025-1.1.5-NEMZ\_KI-2025-00004, and MATE KKP and KKPCs Research Excellence and Flagship Research Groups grants (Hungary); the Council of Science and Industrial Research, India; ICSC -- National Research C enter for High Performance Computing, Big Data and Quantum Computing, FAIR -- Future Artificial Intelligence Research, and CUP I53D23001070006 (Mission 4 Component 1), funded by the NextGenerationEU program, the Italian Ministry of University and Research (MUR) under Bando PRIN 2022 -- CUP I53C24002390006, PRIN PRIMULA 2022RBYK7T (Italy); the Latvian Council of Science; the Ministry of Science and Higher Education, project no. 2022/WK/14, and the National Science C enter, contracts Opus 2021/41/B/ST2/01369, 2021/43/B/ST2/01552, 2023/49/B/ST2/03273, and the NAWA contract BPN/PPO/2021/1/00011 (Poland); the Funda\c{c}\~ao para a Ci\^encia e a Tecnologia (Portugal); the National Priorities Research Program by Qatar National Research Fund; MICIU/AEI/10.13039/501100011033, ERDF/EU, ``European Union NextGenerationEU/PRTR", projects PID2022-142604OB-C21, PID2022-139519OB-C21, PID2023-147706NB-I00, PID2023-148896NB-I00, PID2023-146983NB-I00, PID2023-147115NB-I00, PID2023-148418NB-C41, PID2023-148418NB-C42, PID2023-148418NB-C43, PID2023-148418NB-C44, PID2024-158190NB-C22, RYC2021-033305-I, RYC2024-048719-I, CNS2023-144781, CNS2024-154769 and Plan de Ciencia, Tecnolog{\'i}a e Innovaci{\'o}n de Asturias, Spain; the Chulalongkorn Academic into Its 2nd Century Project Advancement Project, the National Science, Research and Innovation Fund program IND\_FF\_68\_369\_2300\_097, and the Program Management Unit for Human Resources \& Institutional Development, Research and Innovation, grant B39G680009 (Thailand); the Eric \& Wendy Schmidt Fund for Strategic Innovation through the CERN Next Generation Triggers project under grant agreement number SIF-2023-004; the Kavli Foundation; the Nvidia Corporation; the SuperMicro Corporation; the Welch Foundation, contract C-1845; and the Weston Havens Foundation (USA).
\end{acknowledgments}\section*{Data availability} Release and preservation of data used by the CMS Collaboration as the basis for publications is guided by the  \href{https://doi.org/10.7483/OPENDATA.CMS.1BNU.8V1W}{CMS data preservation, re-use and open access policy}.

\bibliography{auto_generated}

\cleardoublepage \appendix\section{The CMS Collaboration \label{app:collab}}\begin{sloppypar}\hyphenpenalty=5000\widowpenalty=500\clubpenalty=5000\input{B2G-22-004-public-authorlist.tex}\end{sloppypar}
\end{document}

%% file: B2G-22-004-public-authorlist.tex
\cmsinstitute{Yerevan Physics Institute, Yerevan, Armenia}
{\tolerance=6000
A.~Hayrapetyan, V.~Makarenko\cmsorcid{0000-0002-8406-8605}, A.~Tumasyan\cmsAuthorMark{1}\cmsorcid{0009-0000-0684-6742}
\par}
\cmsinstitute{Institut f\"{u}r Hochenergiephysik, Vienna, Austria}
{\tolerance=6000
W.~Adam\cmsorcid{0000-0001-9099-4341}, J.W.~Andrejkovic, L.~Benato\cmsorcid{0000-0001-5135-7489}, T.~Bergauer\cmsorcid{0000-0002-5786-0293}, M.~Dragicevic\cmsorcid{0000-0003-1967-6783}, C.~Giordano\cmsorcid{0000-0001-6317-2481}, P.S.~Hussain\cmsorcid{0000-0002-4825-5278}, M.~Jeitler\cmsAuthorMark{2}\cmsorcid{0000-0002-5141-9560}, N.~Krammer\cmsorcid{0000-0002-0548-0985}, A.~Li\cmsorcid{0000-0002-4547-116X}, D.~Liko\cmsorcid{0000-0002-3380-473X}, M.~Matthewman, I.~Mikulec\cmsorcid{0000-0003-0385-2746}, J.~Schieck\cmsAuthorMark{2}\cmsorcid{0000-0002-1058-8093}, R.~Sch\"{o}fbeck\cmsAuthorMark{2}\cmsorcid{0000-0002-2332-8784}, D.~Schwarz\cmsorcid{0000-0002-3821-7331}, M.~Shooshtari\cmsorcid{0009-0004-8882-4887}, M.~Sonawane\cmsorcid{0000-0003-0510-7010}, W.~Waltenberger\cmsorcid{0000-0002-6215-7228}, C.-E.~Wulz\cmsAuthorMark{2}\cmsorcid{0000-0001-9226-5812}
% ...
\par}
\vskip\cmsinstskip
\dag:~Deceased\\
$^{1}$Also at Yerevan State University, Yerevan, Armenia\\
$^{2}$Also at TU Wien, Vienna, Austria\\
% ...
